\documentclass[twocolumn,english]{revtex4}
\usepackage[T1]{fontenc}
\usepackage[latin1]{inputenc}
\usepackage{float}
\usepackage{amsmath}
\usepackage{graphicx}
\usepackage{amssymb}

\makeatletter

\providecommand{\tabularnewline}{\\}

\usepackage{mathptmx}
\usepackage[scaled=0.9]{helvet} 

\usepackage{babel}
\makeatother
\begin{document}

\title{Numerical Simulation of Two Dimensional Electron Transport in a Circularly
Symmetric Cylindrical Nanostructure using Wigner Function Methods}

\author{Greg Recine, Bernard Rosen, Hong-Liang Cui}

\affiliation{Applied Electronics Laboratory\\
Stevens Institute of Technology\\
Hoboken, NJ 07030}

\homepage{http://ael.phy.stevens-tech.edu}

\email{grecine@stevens.edu}

\begin{abstract}
We have constructed a lattice Wigner-Weyl code that generalizes the
Buot-Jensen algorithm to the calculation of electron transport in
two-dimensional circular-cylindrically symmetric structures, where
the Wigner function equation is solved self-consistently with the
Poisson equation. Almost all of the numerical simulations to date
have dealt with the restriction of the problem to one dimensional
transport. In real devices, electrons are not confined to a single
dimension and the coulombic potential is fully present and felt in
three dimensions. We show the derivation of the 2D equation in cylindrical
coordinates as well as approximations employed in the calculation
of the four-dimensional convolution integral of the Wigner function
and the potential. We work under the assumption that longitudinal
transport is more dominant than radial transport and employ parallel
processing techniques. The total transport is calculated in two steps:
(1) transport the particles in the longitudinal direction in each
shell separately, then (2) each shell exchanges particles with its
nearest neighbor. Most of this work is concerned with the former step:
A 1D space and 2D momentum transport problem. Time evolution simulations
based on these method are presented for three different cases. Each
case lead to numerical results consistent with expectations. Discussions
of future improvements are discussed.
\end{abstract}
\maketitle

\section{Introduction}

\begin{center}%
\begin{figure}
\begin{center}\begin{tabular}{cc}
\includegraphics[%
  width=0.35\columnwidth,
  keepaspectratio]{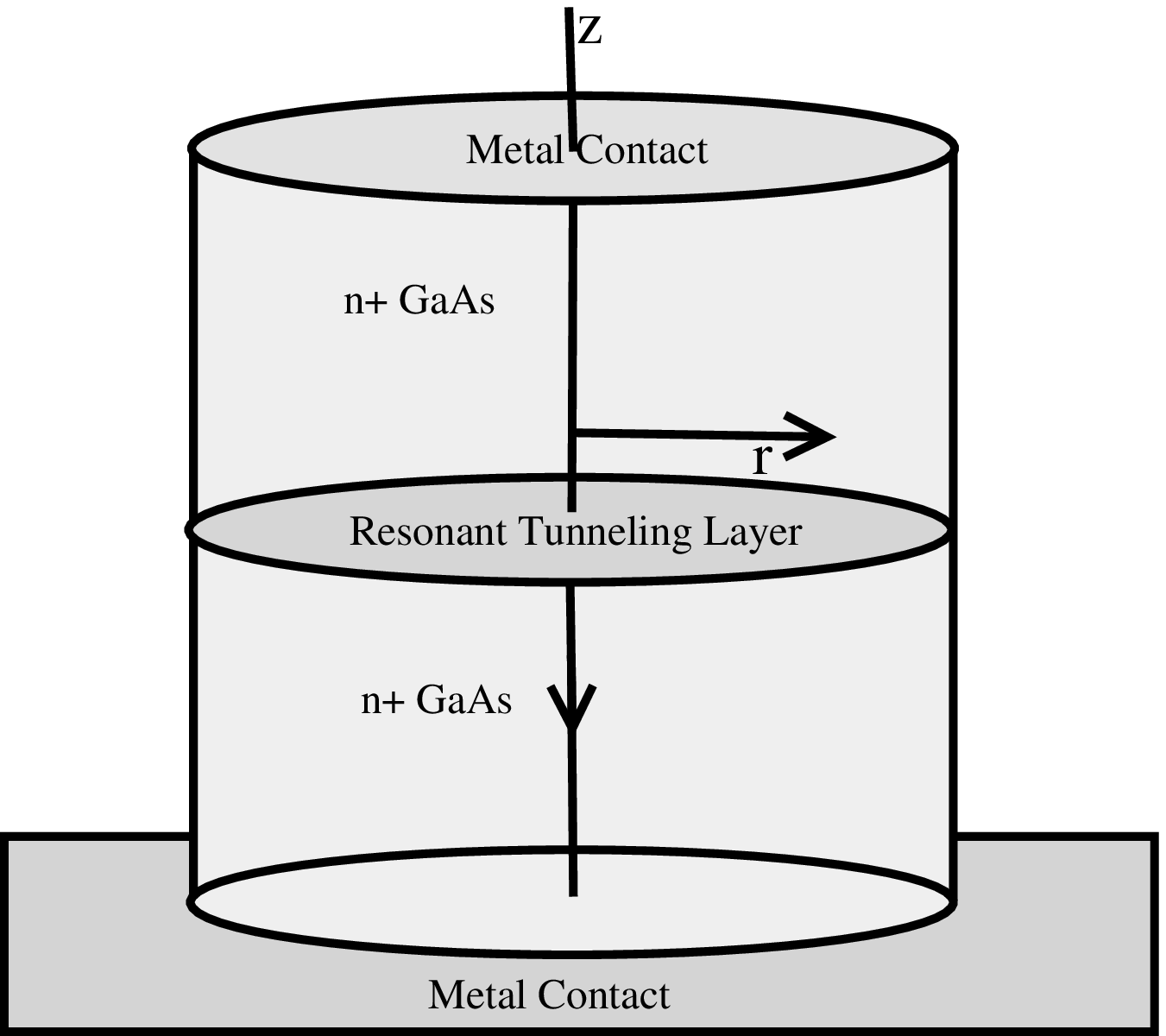}&
\includegraphics[%
  width=0.30\columnwidth,
  keepaspectratio]{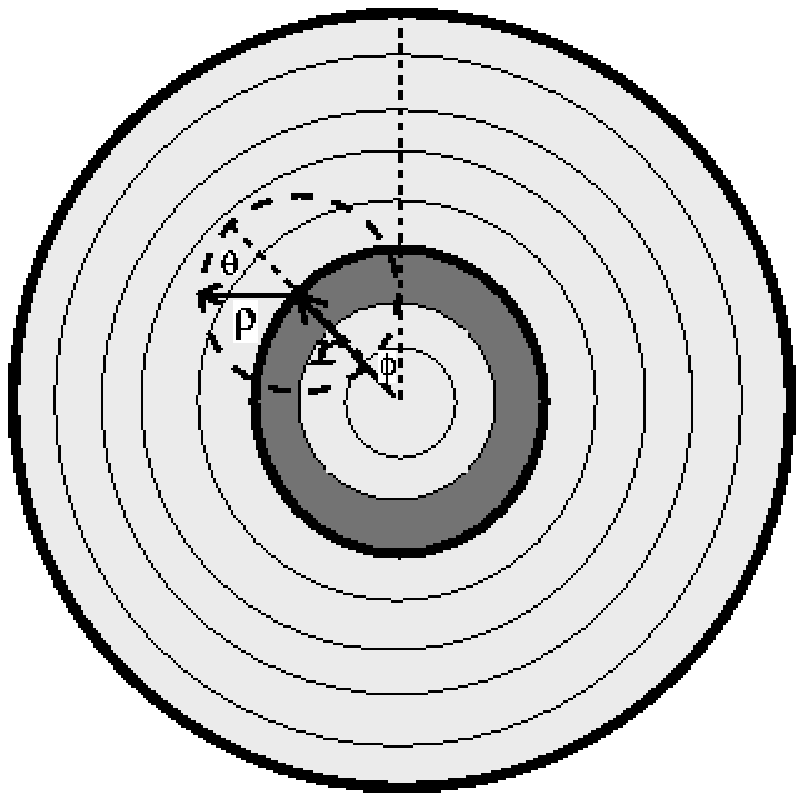}\tabularnewline
(a) Side View&
(b) Top View\tabularnewline
\end{tabular}\end{center}

\caption{\label{cap:Cylindrical-RTD}Cylindrical RTD}
\end{figure}
\end{center}

Electron transport in a resonant tunneling structure (RTS) has been
studied in detail over the past decade \cite{Buot-Jensen-PRB-1990,Goldman-Tsui,Frensley-PRB-1987,Frensley-RMP-1990,RTD-Book,Zhao-Origin_Hyst-plateau}.
Almost all of the numerical simulations have dealt with the restriction
of the problem to one dimensional transport. Even though much progress
has been made using the 1D theory, in real devices electrons are not
confined to transport in a single dimension and the coulombic potential
is fully present and felt in three dimensions. Here we present a method
for numerical simulation of electronic transport through a cylindrical
device that possesses azimuthal symmetry. Figure \ref{cap:Cylindrical-RTD}(a)
shows a schematic representation of such a device. 

We work under the assumption that longitudinal transport is more dominant
than radial transport. The total transport is calculated in two steps:
(1) transport the particles in the longitudinal direction in each
shell separately, then (2) each shell exchanges particles with its
nearest neighbor. During a given time step the particles are advanced
longitudinally through the device, as in a 1D problem, but with the
inclusion of radial momentum. This changes the form of the potential
and interaction terms of the familiar 1D Wigner function (transport)
equation (WFE). Since the latter step is computationally simple, most
of this work is concerned with the former step: A 1D space and 2D
momentum transport problem (1x+2k). In this paper, the WFE is solved
self-consistently with the Poisson equation.

In order to perform the numerical simulation, parallel programming
techniques are used. A simplest way to attack this problem to slice
up the device into cylindrically concentric shells as shown in figure
\ref{cap:Cylindrical-RTD}(b). The two above steps now become: (1)
each processor (shell) calculates the (1x+2k) transport problem, then
(2) each processor exchanges particle information with its nearest
neighbor.

This papers is organized as follows: The first three sections detail
the derivations: Section \ref{sec:Derivation} the 2D WFE in cylindrical
coordinates (assuming azimuthal symmetry), and sections \ref{sec:Discretization}
and \ref{sec:2D-Matrix-Setup} the discretization. Section \ref{sec:Methods-of-Solution}
discusses different methods of solving the 1x+2k problem regarding
the limitations of today's computational resources. This includes
a re-derivation of the potential term in the WFE and a splitting of
the potential into static (conduction band edge) and changing (self-consistent)
potentials. Finally, our concluding remarks are in section \ref{sec:Conclusions}.

\section{{\normalsize Derivation\label{sec:Derivation}}}

The 3D form of the Wigner function equation (WFE), can be written
as (scattering will be added in later){\small \begin{gather}
\frac{df(\mathbf{q},\mathbf{k})}{dt}=-\frac{\hbar\mathbf{k}}{m^{*}}\cdot\bigtriangledown_{\mathbf{q}}f(\mathbf{q},\mathbf{k})\nonumber \\
-\frac{\textrm{i}}{\left(2\pi\right)^{3}\hbar}\int d\mathbf{k}'\int2d\mathbf{y}e^{-2\textrm{i}\left(\mathbf{\mathbf{k}-k'}\right)\cdot\mathbf{y}}\left\{ V\left(\mathbf{q}+\mathbf{y}\right)-V\left(\mathbf{q}-\mathbf{y}\right)\right\} f(\mathbf{q},\mathbf{k'}).\label{eq:WFE-3D2}\end{gather}
}The derivation of the WFE was in the same vein as Frensley\cite{Frensley-RMP-1990}
(but kept in full 3D form here), and, as he has noted, has been derived
while assuming no boundaries exist. Because of this, the integral
limits have been omitted in the above and will be discussed later.

First, let us rewrite this equation in cylindrical coordinates by
making the substitutions: $\mathbf{q}\rightarrow r,z,\phi\mathbf{\,;\, k}\rightarrow k_{r},k_{z},\chi_{\phi}\,;\,\mathbf{y}\rightarrow\rho,\zeta,\theta$.
It is important to note that while $\mathbf{q}$ represents the spatial
(i.e.: center of mass) coordinate with respect to the origin on the
cylindrical axis, $\mathbf{y}$ represents the spatial coordinate
with respect to the origin at an arbitrary point within the cylinder,
not necessarily on the axis. Figure \ref{cap:Cylindrical-RTD}(b)
illustrates the geometry of these two variables. Equation \ref{eq:WFE-3D2}
now becomes:{\small \begin{gather}
\frac{df(r,z,\phi,k_{z},k_{r},\chi_{\phi})}{dt}=\nonumber \\
-\frac{\hbar}{m^{*}}\left[k_{r}\frac{\partial}{\partial r}+\frac{\chi_{\phi}}{r}\frac{\partial}{\partial\phi}+k_{z}\frac{\partial}{\partial z}\right]f(r,z,\phi,k_{z},k_{r},\chi_{\phi})-\nonumber \\
\frac{2\textrm{i}}{\left(2\pi\right)^{3}\hbar}\int dk'_{z}\int dk'_{r}\int_{0}^{2\pi}\left|k'_{r}\right|d\chi'_{\phi}\int d\zeta\int d\rho\int_{0}^{2\pi}\left|\rho\right|d\theta\times\nonumber \\
e^{-2\textrm{i}\left[\left(k_{z}-k'_{z}\right)\zeta+\left(k_{r}\cos\chi_{\phi}-k'_{r}\cos\chi'_{\phi}\right)\rho\right]}\times\nonumber \\
\left\{ V\left(r+\rho,z+\zeta,\phi+\theta\right)-V\left(r-\rho,z-\zeta,\phi-\theta\right)\right\} f(r,z,\phi,k'_{z},k'_{r},\chi'_{\phi})\end{gather}
}Considering a 2D problem in cylindrical coordinates with azimuthal
symmetry, the first thing to notice is that even though the potential
is symmetric in $\phi$ ($V\left(r,z,\phi\right)\rightarrow V\left(r,z\right)$),
the potential difference \emph{is dependent on angle}, $V\left(r\pm\rho,z\pm\zeta,\phi\pm\theta\right)\rightarrow V\left(r\pm\rho,z\pm\zeta,\pm\theta\right)$.
Next, integrate out the remaining azimuthal spatial and momentum components.
Using{\small \begin{equation}
\int_{0}^{2\pi}\int_{0}^{2\pi}f(r,z,\phi,k_{z},k_{r},\chi_{\phi})\left|r\right|d\phi\left|k_{r}\right|d\chi_{\phi}=(2\pi)^{2}\left|r\right|\left|k_{r}\right|f(r,z,k_{z},k_{r}),\end{equation}
}one obtains\begin{gather}
\frac{df(r,z,k_{z},k_{r})}{dt}=-\frac{\hbar}{m^{*}}\left[k_{r}\frac{\partial}{\partial r}+k_{z}\frac{\partial}{\partial z}\right]f(r,z,k_{z},k_{r})-\nonumber \\
\frac{2\textrm{i}}{\left(2\pi\right)^{3}\hbar}\int dk'_{z}\int dk'_{r}\int_{0}^{2\pi}\left|k'_{r}\right|d\chi'_{\phi}\int d\zeta\int d\rho\int_{0}^{2\pi}\left|\rho\right|d\theta\times\nonumber \\
\int_{0}^{2\pi}d\chi_{\phi}e^{-2\textrm{i}\left[\left(k_{z}-k'_{z}\right)\zeta+\left(k_{r}\cos\chi_{\phi}-k'_{r}\cos\chi'_{\phi}\right)\rho\right]}\times\nonumber \\
\left\{ V\left(z+\zeta,r+\rho,\theta\right)-V\left(z-\zeta,r-\rho,-\theta\right)\right\} f(r,z,k'_{z},k'_{r},\chi'_{\phi}).\label{eq:full glory 2D}\end{gather}
The drift term is relatively simple, but the potential term is a bit
completed. Rewriting the potential term as\begin{gather}
-\frac{2\textrm{i}}{\left(2\pi\right)^{3}\hbar}\int\left|k'_{r}\right|dk'_{r}\int dk'_{z}\int d\zeta e^{-2\textrm{i}\left(k_{z}-k'_{z}\right)\zeta}\times\nonumber \\
\int\left|\rho\right|d\rho\,\mathcal{J}(\rho,k_{r})\,\mathcal{V}(z,\zeta,r,\rho)\,\mathcal{F}(z,r,\rho,k'_{z},k'_{r}),\end{gather}
using the following definitions{\small \begin{equation}
\mathcal{J}(\rho,k_{r})=\int_{0}^{2\pi}d\chi_{\phi}e^{-2\textrm{i}k_{r}\cos\chi_{\phi}\rho}\label{eq:U(r',kr)}\end{equation}
\begin{equation}
\mathcal{V}(z,\zeta,r,\rho)=\int_{0}^{2\pi}d\theta\left\{ V\left(z+\zeta,r+\rho,\theta\right)-V\left(z-\zeta,r-\rho,-\theta\right)\right\} \label{eq:V(z,z',r,r')}\end{equation}
\begin{equation}
\mathcal{F}(z,r,\rho,k'_{z},k'_{r})=\int_{0}^{2\pi}d\chi'_{\phi}e^{+2\textrm{i}k'_{r}\cos\chi'_{\phi}\rho}f(r,z,k'_{z},k'_{r},\chi'_{\phi}),\label{eq:F(z,r,r',kz',kr')}\end{equation}
}there are some reductions in complexity. Equation \ref{eq:U(r',kr)}
is just the definition of the Bessel function $2\pi J_{0}(2k_{r}\rho)$.
Equation \ref{eq:V(z,z',r,r')} is evaluated at a given $\rho,\theta$
by noting that $\rho(\theta)=\rho\cos\theta$, leaving equation \ref{eq:F(z,r,r',kz',kr')}
to be dealt with. By expanding the exponential in terms of Bessel
functions of $\chi'_{\phi}$, the integral becomes{\small \begin{eqnarray}
\mathcal{F}(z,r,\rho,k'_{z},k'_{r}) & = & \int_{0}^{2\pi}d\chi'_{\phi}2\pi\sum_{m,m'}J_{m'}(2k'_{r}\rho)f_{m}(r,z,k'_{z},k'_{r})e^{i(m+m')\chi'_{\phi}}\nonumber \\
 & = & 2\pi\sum_{m}J_{m}(2k'_{r}\rho)f_{m}(r,z,k'_{z},k'_{r}).\end{eqnarray}
}Now there is an infinite series in $m$, but only the term $m=0$
must be counted. This is because it represents the azimuthally independent
functions, which, as dictated by the current problem, is the form
that the Wigner distribution function, $f$, should take.

So, finally, the complete 2D form of the WFE in azimuthally independent
cylindrical coordinates is\begin{gather}
\frac{df(r,z,k_{z},k_{r})}{dt}=-\frac{\hbar}{m^{*}}\left[k_{r}\frac{\partial}{\partial r}+k_{z}\frac{\partial}{\partial z}\right]f(r,z,k_{z},k_{r})\nonumber \\
+\frac{1}{\pi\hbar}\int dk'_{z}\int\left|\rho\right|d\rho\mathcal{U}(r,\rho,z,k_{r},k_{z}-k'_{z})\mathcal{F}(r,\rho,z,k'_{z}),\label{eq:WFE-2D_ptI}\end{gather}
where (notice the terms $\mathcal{F}$, $\mathcal{U}$ and $\mathcal{V}$
have been redefined from what was written above){\small \begin{equation}
\mathcal{F}(r,\rho,z,k_{z})=\int\left|k'_{r}\right|dk'_{r}J_{0}(2k'_{r}\rho)f(r,z,k_{z},k_{r}')\label{eq:F}\end{equation}
\begin{equation}
\mathcal{U}(r,\rho,z,k_{z},k_{r})=\int d\zeta\sin(2k_{z}\zeta)J_{0}(2k_{r}\rho)\mathcal{V}(z,\zeta,r,\rho)\end{equation}
\begin{equation}
\mathcal{V}(z,\zeta,r,\rho)=\int_{0}^{2\pi}d\theta\left\{ V\left(z+\zeta,r+\rho\cos\theta\right)-V\left(z-\zeta,r-\rho\cos\theta\right)\right\} \end{equation}
}The integral limits are ${\displaystyle \int}_{-k_{z}^{max}}^{+k_{z}^{max}}dk'_{z}$,
${\displaystyle \int}_{-k_{r}^{max}}^{+k_{r}^{max}}dk'_{r}$, ${\displaystyle \int}_{0}^{L/2}d\zeta$,
and${\displaystyle \int}_{0}^{R/2}\rho d\rho$ for a cylindrical system
of length $L$ and radius $R$, recognizing that $r>0$ always. It
is worthwhile to note here that the since the momentum variable comes
from the Fourier transform $\int d\mathbf{r}e^{-\textrm{i}\mathbf{k\cdot r}}$,
the value of $k^{max}$ is determined by the spatial length of the
box. This will become important when the problem is discretized. 

As done by many others\cite{Buot-Jensen-PRB-1990,Frensley-RMP-1990},
scattering is included simply by the addition of a relaxation time
approximation, given as\begin{gather}
\left.\frac{df(r,z,k_{z},k_{r})}{dt}\right|_{coll}=\nonumber \\
\frac{1}{\tau}\left(f_{0}(r,z,k_{z},k_{r})\frac{\int\left|k'_{r}\right|dk'_{r}\int dk'_{z}\, f(r,z,k'_{z},k'_{r})}{\int\left|k'_{r}\right|dk'_{r}\int dk'_{z}\, f_{0}(r,z,k'_{z},k'_{r})}-f(r,z,k_{z},k_{r})\right),\end{gather}
where the relaxation time, $\tau$ is computed from the material parameters
describing scattering due to: ionized impurities and longitudinal,
piezoelectric, acoustic \& optical phonons.

\section{Discretization\label{sec:Discretization}}

The WFE in discretized (matrix) form is written as $\frac{d\mathbf{f}}{dt}=\left(\mathbf{T}+\mathbf{U}+\mathbf{S}\right)\mathbf{f}-\mathbf{B}$,
where $\mathbf{T}$ represents the drift (kinetic) operator, $\mathbf{U}$
the potential operator, $\mathbf{S}$ the scattering (interaction)
operator, $\mathbf{f}$ the Wigner function and $\mathbf{B}$ the
boundary conditions arising from the drift term. In this section,
we will present the details of our discretization of the equations
derived above.

\subsection{Variables, Operators and Functions}

The space and momentum variables are discretized as{\small \begin{eqnarray}
z(i) & = & \frac{1}{2}\left(2i-1\right)\Delta z\,\,\,,\,\,\, i=1..N_{z}\,\,\,,\,\,\,\Delta z=L/N_{z}\\
r(n) & = & \frac{1}{2}\left(2n-1\right)\Delta r\,\,\,,\,\,\, n=1..N_{r}\,\,\,,\,\,\,\Delta r=R/N_{r}\\
\phi(m) & = & \left(m-1\right)\Delta\phi\,\,\,,\,\,\, m=1..N_{\phi}\,\,\,,\,\,\,\Delta\phi=2\pi/N_{\phi}\\
k_{z}(j) & = & \frac{1}{2}\left(2j-N_{k_{z}}-1\right)\Delta k_{z}\,,\, j=1..N_{k_{z}}\,,\,\Delta k_{z}=\pi/\Delta zN_{k_{z}}\\
k_{r}(l) & = & \frac{1}{2}\left(2k-N_{k_{r}}-1\right)\Delta k_{r}\,,\, k=1..N_{k_{r}}\,,\,\Delta k_{r}=\pi/\Delta rN_{k_{r}}\end{eqnarray}
}and the functions $f$, $\mathcal{V}$, and $\mathcal{U}$, are discretized
as\begin{eqnarray}
f(r,z,k_{z},k_{r}) & \rightarrow & f(n,i,j,l)\\
V(r-r',z-z') & \rightarrow & V(n-n',i-i')\\
\mathcal{U}(r,r',z,k_{z},k_{r}) & \rightarrow & \mathcal{U}(n,n',i,j,l)\\
\mathcal{V}(z,z',r,r') & \rightarrow & \mathcal{V}(i,i',n,n')\\
\mathcal{F}(r,r',z,k_{z}) & \rightarrow & \mathcal{F}(n,n',i,j).\end{eqnarray}
We note here that since $\rho$ and $\zeta$ are on the same grid
as $r$ and $z$, they will be denoted as $r'$ and $z'$, respectively.

\subsection{2D Poisson Equation}

We use a Fourier transform method to solve the 2D Poisson equation.
The Fourier transform of the charge density with respect to $z$,
$\overline{\rho_{e}}$, is calculated using a fast Fourier sin transform
(sinFFT). Each shell exchanges $\overline{\rho_{e}}$ to every other
shell so that each shell knows $\overline{\rho_{e}}(r)$, the electron
density of the entire cylinder. Using a standard tridiagonal solver,
the sinFFT of the 2D potential, $\overline{\phi}(r)$ is calculated,
then transformed back (via a sinFFT) to the full 2D potential, $\phi(r,z)$.

\subsection{Drift \& Boundary Conditions Terms\label{sub:Drift-&-Boundary}}

We will separate the drift (or kinetic) term into longitudinal and
radial components and treat each one slightly differently\begin{equation}
\left[\mathbf{T}\cdot\mathbf{f}\right](r,z,k_{z},k_{r})=-\frac{\hbar}{m^{*}}\left[k_{r}\frac{\partial}{\partial r}+k_{z}\frac{\partial}{\partial z}\right]f(r,z,k_{z},k_{r})\end{equation}
which is broken up into \begin{eqnarray}
\left[\mathbf{T_{z}}\cdot\mathbf{f}\right](r,z,k_{z},k_{r}) & = & -\frac{\hbar k_{z}}{m^{*}}\frac{\partial}{\partial z}f(r,z,k_{z},k_{r})\label{eq:z-drift}\\
\left[\mathbf{T_{r}}\cdot\mathbf{f}\right](r,z,k_{z},k_{r}) & = & -\frac{\hbar k_{r}}{m^{*}}\frac{\partial}{\partial r}f(r,z,k_{z},k_{r})\label{eq:r-drift}\end{eqnarray}
First the longitudinal drift term, equation \ref{eq:z-drift}, is
computed using a second order {}``upwind/downwind'' differencing
scheme:\begin{equation}
\frac{df(x)}{dx}\simeq\mp\frac{1}{2\bigtriangleup x}\left[3f(x)-4f(x\pm\bigtriangleup x)+f(x\pm2\bigtriangleup x)\right].\end{equation}
For $k_{z}<0$, the upwind scheme is used, and for $k_{z}>0$, the
downwind scheme is used, giving{\small \begin{gather}
k_{z}\lessgtr0:\,\,\frac{\partial}{\partial z}f(r,z,k_{z},k_{r})\rightarrow\nonumber \\
\mp\frac{1}{2\bigtriangleup z}\left[3f(r,z,k_{z},k_{r})-4f(r,z\pm\bigtriangleup z,k_{z},k_{r})+f(r,z\pm2\bigtriangleup z,k_{z},k_{r})\right].\end{gather}
}Which, when discretized, gives{\small \begin{gather}
j\lesseqgtr\frac{N_{k_{z}}}{2}:\,\,\left[\mathbf{T_{z}}\cdot\mathbf{f}\right](n,i,j,l)\rightarrow\pm\frac{\hbar\bigtriangleup k_{z}}{8m^{*}\bigtriangleup z}\times\nonumber \\
\left(2j-N_{k_{z}}-1\right)\left[3f(n,i,j,l)-4f(n,i\pm1,j,l)+f(n,i\pm2,j,l)\right].\end{gather}
}When the second order differencing scheme gets to the boundary, it
is advantageous to only have the function extend one unit distance
into the boundary. For the upwind scheme, this occurs at $i=Nz-1$
and $i=Nz$, and $i=1$ and $i=2$ at for the downwind scheme. (the
$n,l$ indexes will be dropped since there is no dependence on them).
When $i=1,N_{z}$ a first order upwind/downwind differencing scheme
was employed $\left(\frac{df(x)}{dx}\simeq\mp\frac{f(x)-f(x\pm1)}{\bigtriangleup x}\right)$
in order to preserve the continuity of the derivative. By saying that
the distribution function past the boundaries have a constant value,
$f(i=N_{z}+1,j)=f(i=0,j)=f_{Fermi}(j)$ (the two dimensional Fermi
distribrution), these positions become constant longitudinal boundary
conditions, $\mathbf{B_{z}}(i,j)$, defined as:\begin{eqnarray}
 &  & j>\frac{N_{k_{z}}}{2}:\nonumber \\
\mathbf{B}_{z}(i=1,j) & = & +2C_{j}f_{Fermi}(j)\label{eq:BC-begin}\\
\mathbf{B_{z}}(2,j) & = & -C_{j}f_{Fermi}(j)\\
 &  & \cdots\cdots\cdots\nonumber \\
 &  & j\leq\frac{N_{k_{z}}}{2}:\nonumber \\
\mathbf{B_{z}}(N_{z}-1,j) & = & +C_{j}f_{Fermi}(j)\\
\mathbf{B_{z}}(N_{z},j) & = & -2C_{j}f_{Fermi}(j).\label{eq:BC-end}\end{eqnarray}
Where $C_{j}=\frac{\hbar\bigtriangleup k_{z}}{4m^{*}\bigtriangleup z}\left(2j-N_{k_{z}}-1\right)$.
This makes the final form of the longitudinal drift term, $\left[\mathbf{T_{z}}\cdot\mathbf{f}\right](i,j)$,
as:{\small \begin{eqnarray}
 &  & j>\frac{N_{k_{z}}}{2}:\nonumber \\
\left[\mathbf{T_{z}}\cdot\mathbf{f}\right](i=1,j) & = & -2C_{j}\left[2f(i=1,j)\right]\label{eq:Tzf-beg}\\
\left[\mathbf{T_{z}}\cdot\mathbf{f}\right](2,j) & = & -C_{j}\left[3f(i=2,j)-4f(i=1,j)\right]\\
\left[\mathbf{T_{z}}\cdot\mathbf{f}\right](i,j) & = & -C_{j}\left[3f(i,j)-4f(i,j)+f(i,j)\right]\label{eq:Tzf-pos}\\
 &  & \cdots\cdots\cdots\cdots\nonumber \\
 &  & j\leq\frac{N_{k_{z}}}{2}:\nonumber \\
\left[\mathbf{T_{z}}\cdot\mathbf{f}\right](i,j) & = & C_{j}\left[3f(i,j)-4f(i,j)+f(i,j)\right]\label{eq:Tzf-neg}\\
\left[\mathbf{T_{z}}\cdot\mathbf{f}\right](N_{z}-1,j) & = & C_{j}\left[3f(i=N_{z}-1,j)-4f(i=N_{z},j)\right]\\
\left[\mathbf{T_{z}}\cdot\mathbf{f}\right](N_{z},j) & = & 2C_{j}\left[2f(i=N_{z},j)\right].\label{eq:Tzf-end}\end{eqnarray}
}Equations \ref{eq:BC-begin} to \ref{eq:BC-end} completely define
the discretized longitudinal boundary conditions, while equations
\ref{eq:Tzf-beg} to \ref{eq:Tzf-pos} completely define the longitudinal
drift for positive momenta and equations \ref{eq:Tzf-neg} to \ref{eq:Tzf-end}
completely define the longitudinal drift for negative momenta.

Next, the radial drift term, $\mathbf{T_{r}}\cdot\mathbf{f}$, is
computed using a first order differencing scheme since having each
radial shell talking only to its nearest neighbor is needed when this
algorithm is parallelized. For a shell that has neighbors on both
sides, a central differencing scheme (CDS) is used. For the innermost
and outermost shell, a forward/backwards differencing scheme (FBDS)
is employed (the indexes $i,j$ are omitted since there is no dependence
on them) ($C=\left(\frac{\hbar k_{r}}{m^{*}\bigtriangleup r}\right)$):{\small \begin{eqnarray}
\left[\mathbf{T_{r}}\cdot\mathbf{f}\right](n=1,l) & \rightarrow & -C\left[f(n=1,l)-f(n=2,l)\right]\label{eq:radial_drift_in}\\
\left[\mathbf{T_{r}}\cdot\mathbf{f}\right](n,l) & \rightarrow & -\frac{C}{2}\left[f(n+1,l)-f(n-1,l)\right]\label{eq:radial_drift}\\
\left[\mathbf{T_{r}}\cdot\mathbf{f}\right](n=N_{r},l) & \rightarrow & C\left[f(n=N_{r},l)-f(n=N_{r}-1,l)\right].\label{eq:radial_drift_out}\end{eqnarray}
}This form dictates that some tricks must be used at the innermost
and outermost rings. For the innermost shell, it can be imagined that
any particle possessing negative momenta (traveling inwards) will
pass through the middle and then posses positive momenta (traveling
outwards). This demands a {}``particle mirror'' at the origin by,
basically, saying that all values of $f\left(1,l\leq\frac{N_{k_{r}}}{2}\right)$
at time $t$, will be added to $f\left(1,l>\frac{N_{k_{r}}}{2}\right)$
at time $t+\Delta t$. For the outermost shell, all values of $f\left(N_{r},l>\frac{N_{k_{r}}}{2}\right)$
at time $t$, will be subtracted from $f\left(1,l>\frac{N_{k_{r}}}{2}\right)$
at time $t+\Delta t$ and, depending on the chosen external conditions,
given values of $f\left(N_{r},l\leq\frac{N_{k_{r}}}{2}\right)$ will
be added at each time step. This is expressed as radial boundary conditions
as ($C_{l}=\frac{\hbar\bigtriangleup k_{r}}{m^{*}\bigtriangleup r}\left(2l-N_{k_{r}}-1\right)$):{\small \begin{eqnarray}
\mathbf{B_{r}}(n=1,l\leq\frac{N_{k_{r}}}{2}) & = & -C_{l}\left[-f(n=1,l\leq\frac{N_{k_{r}}}{2})\right]\label{eq:BC-inner-1}\\
\mathbf{B_{r}}(n=1,l>\frac{N_{k_{r}}}{2}) & = & -C_{l}\left[+f(n=1,l\leq\frac{N_{k_{r}}}{2})\right]\label{eq:BC-inner-2}\\
\mathbf{B_{r}}(n=N_{r},l>\frac{N_{k_{r}}}{2}) & = & +C_{l}\left[-f(n=N_{r},l>\frac{N_{k_{r}}}{2})\right].\label{eq:BC-outer-exit}\\
\mathbf{B_{r}}(n=N_{r},l\leq\frac{N_{k_{r}}}{2}) & = & +C_{l}\left[+f_{Given}(l)\right].\label{eq:BC-outer-in}\end{eqnarray}
}{\small \par}

\subsection{Potential Term}

The potential term in equation \ref{eq:WFE-2D_ptI} is written in
operator form as: {\small \begin{gather}
\left[\mathbf{U\cdot f}\right](r,z,k_{z},k_{r})\equiv\nonumber \\
+\frac{1}{\pi\hbar}\int_{-k_{z}^{max}}^{+k_{z}^{max}}dk'_{z}\int_{0}^{R/2}\left|r'\right|dr'\mathcal{U}(r,r',z,k_{z}-k'_{z},k_{r})\mathcal{F}(r,r',z,k'_{z})\label{eq:Uf-continuous}\end{gather}
}where,{\small \begin{eqnarray}
\mathcal{V}(z,z',r,r') & = & \int_{0}^{2\pi}d\phi'\left\{ V\left(z+z',r+r'\cos\phi'\right)-\right.\label{eq:V-continuous}\\
 &  & \left.-V\left(z-z',r-r'\cos\phi'\right)\right\} \nonumber \\
\mathcal{F}(r,r',z,k_{z}) & = & {\displaystyle \int}_{-k_{r}^{max}}^{+k_{r}^{max}}\left|k'_{r}\right|dk'_{r}J_{0}(2k'_{r}r')f(r,z,k_{z},k_{r}')\label{eq:F-continuous}\\
\mathcal{U}(r,r',z,k_{z},k_{r}) & = & {\displaystyle \int}_{0}^{L/2}dz'\sin(2k_{z}z')J_{0}(2k_{r}r')\mathcal{V}(z,z',r,r')\label{eq:U-continuous}\end{eqnarray}
}For a given longitudinal position and radius, $(z,r)$, one sweeps
through disks of constant $z'$. For each disk of constant $z'$,
the potential difference contribution, $V\left(r+r'\cos\phi'\right)-V\left(r-r'\cos\phi'\right)$,
between all points on the disk on $(z,r)$ is calculated (see figure
\ref{cap:Cylindrical-RTD}(b)). The result is that the effect of the
potential of the entire cylinder on a given point $(z,r)$ operates
on the distribution function for that point. The terms $\mathcal{U}$,
$\mathcal{F}$ and $\mathcal{V}$ each present some difficulties,
but they only involve one integral each. It is best to go through
each one separately.

The $\mathcal{V}$ term, when discretized, becomes\begin{equation}
\mathcal{V}(i,i',n,n')=\Delta\phi\sum_{m'=1}^{N_{\phi}}\left\{ V\left(i+i',n+n''\right)-V\left(i-i',n-n''\right)\right\} ,\end{equation}
where\begin{equation}
n''=n'INT\left[\cos\phi(m')\right]=n'INT\left[\cos\left(\left[m'-1\right]\Delta\phi\right)\right],\end{equation}
$INT[x]$ being a function returning the nearest integer to the real
value $x$. 

We rewrite the potential term, by introducing new functions, as{\footnotesize \begin{equation}
\left[\mathbf{U\cdot f}\right](r,z,k_{z},k_{r})=\frac{1}{\pi\hbar}\int_{-k_{z}^{max}}^{+k_{z}^{max}}dk'_{z}{\displaystyle \int}_{-k_{r}^{max}}^{+k_{r}^{max}}dk'_{r}U(r,z,k_{z}-k'_{z},k_{r},k'_{r})f(r,z,k'_{z},k'_{r})\end{equation}
}{\small \begin{equation}
U(r,z,k_{z},k_{r},k'_{r})=\left|k'_{r}\right|\int_{0}^{R/2}dr'\left|r'\right|J_{0}(2k'_{r}r')J_{0}(2k_{r}r')\mathcal{P}(r,r',z,k_{z})\end{equation}
\begin{equation}
\mathcal{P}(r,r',z,k_{z})={\displaystyle \int}_{0}^{L/2}dz'\sin(2k_{z}z')\mathcal{V}(z,z',r,r')\end{equation}
}where $\mathcal{V}(z,z',r,r')$ is defined as above. In discretized
form, these equations become{\footnotesize \begin{eqnarray}
\left[\mathbf{U\cdot f}\right](n,i,j,l) & = & \frac{\pi^{2}}{\hbar N_{k_{z}}N_{k_{r}}^{2}}\sum_{j'}\sum_{l'}U(n,i,j-j',l,l')f(n,i,j',l')\label{eq:2D_Potential_Discrete}\\
U(n,i,j,l,l') & = & \left|(2l'-N_{k_{r}}-1)\right|\sum_{n'}\mathcal{J}(n',l,l')\mathcal{P}(n,n',i,j)\\
\mathcal{J}(n,l,l') & = & \left|(2n'-1)\right|J_{0}\left(\frac{\pi}{N_{k_{r}}}(2l'-N_{k_{r}}-1)(2n-1)\right)\\
 &  & J_{0}\left(\frac{\pi}{N_{k_{r}}}(2l-N_{k_{r}}-1)(2n-1)\right)\nonumber \\
\mathcal{P}(n,n',i,j) & = & \sum_{i'}\sigma(i,j)\mathcal{V}(i,i',n,n')\\
\sigma(i,j) & = & \sin\left(\frac{2\pi}{N_{k_{z}}}j(2i-1)\right)\end{eqnarray}
}In performing all these calculations, it is important to remember
that in current computing platforms, memory is both cheap and nicely
managed so any number of the terms in these arrays can be calculated
once or once per cycle and stored in advance. By employing distributed
computing techniques, a dedicated CPU can spend it's time calculating
these arrays while other parts of the program are running. Parallelization
of this algorithm will be discussed in detail below.

\subsection{Interaction Term}

The scattering term is written using the relaxation time approximation{\footnotesize \begin{gather}
\left.\frac{df(r,z,k_{z},k_{r})}{dt}\right|_{coll}=\nonumber \\
\frac{1}{\tau}\left(\frac{f_{0}(r,z,k_{z},k_{r})}{\int\left|k'_{r}\right|dk'_{r}\int dk'_{z}\, f_{0}(r,z,k'_{z},k'_{r})}\int\left|k'_{r}\right|dk'_{r}\int dk'_{z}\, f(r,z,k'_{z},k'_{r})-f(r,z,k_{z},k_{r})\right)\end{gather}
}where $f_{0}(r,z,k'_{z},k'_{r})$ is the equilibrium WDF. This term,
discretized, becomes{\footnotesize \begin{gather}
\left[\mathbf{S}\cdot\overrightarrow{f}\right](n,i,j,l)=\frac{\beta(n,i,j,l)}{\tau}\sum_{j'=1}^{N_{k_{z}}}\sum_{l'=1}^{N_{k_{r}}}\left|(2l'-N_{k_{r}}-1)\right|f(n,i,j',l')-\frac{1}{\tau}f(n,i,j,l)\nonumber \\
\beta(n,i,j,l)=\frac{f_{0}(n,i,j,l)}{\sum_{j'=1}^{N_{k_{z}}}\sum_{l'=1}^{N_{k_{r}}}\left|(2l'-N_{k_{r}}-1)\right|f_{0}(n,i,j',l')}.\label{eq:2D_Interaction_Discrete}\end{gather}
}{\footnotesize \par}

\section{2D Matrix Setup\label{sec:2D-Matrix-Setup}}

It is important that the matrix $\Omega=\mathbf{T}+\mathbf{U}+\mathbf{S}$
be such that the limitations of modern computer systems are able to
handle the above equations in a reasonable amount of time (days and
weeks as opposed to months). We know of two general methods that are
able to efficiently solve the system of equations generated by the
discretization of the WFE: Matrix inversion techniques and direct
integration techniques (explicit Runge-Kutta-like, implicit BDF/Adams,
etc). For the matrix inversion methods, a necessary condition is that
either the whole matrix can to be stored in RAM or can be broken up
into parts that can be separately stored in RAM and solved for sequentially
(i.e.: block diagonal matrix). The necessary condition for direct
integration techniques is simply that the technique be stable and
fast enough.

Each processor is handling the 1x+2k problem, which implies a large
number or matrix elements/equations. This number is too large to satisfy
the above conditions. In order to remedy this, we try to find a form
of the matrix $\Omega$ that is block diagonal in $k_{r}$, similar
to the 1D problem, but with slightly different terms. We find that
the potential term $\mathbf{U}$ cannot be expressed in such a way,
and present an alternative method of solution, using approximation
techniques, in section \ref{sec:Methods-of-Solution}.

\subsection{Kinetic Matrix (Longitudinal)}

Following the procedure well defined in previous literature\cite{Buot-Jensen-PRB-1990,Frensley-RMP-1990,Zhao-Origin_Hyst-plateau},
we say $[f(i,j)]_{l}$ is written in vector form as$\left[f_{1,1}f_{1,2}f_{1,3}\cdots f_{1,N_{k_{z}}}f_{2,1}\cdots f_{i,j-1}f_{i,j}f_{i,j+1}\cdots f_{N_{z},N_{k_{z}}}\right]_{l}^{T}$
so we can write that, for a given $i$ the longitudinal drift term
can be written in matrix form as (< and > denotes downwind and upwind
differentiation, respectively.)\begin{equation}
\mathcal{T}_{n}^{>}\overrightarrow{f_{i,j,l}}=C\left[\begin{array}{rrrcrr}
2T_{n}\\
 & \ddots\\
 &  & T_{n}\\
 &  &  & 0\\
 &  &  &  & \ddots\\
 &  &  &  &  & 0\end{array}\right]\left[\begin{array}{c}
f_{i,1,l}\\
\vdots\\
f_{i,\frac{N_{k_{z}}}{2},l}\\
f_{i,\frac{N_{k_{z}}}{2}+1,l}\\
\vdots\\
f_{i,N_{k_{z}},l}\end{array}\right]_{l},\end{equation}
and\begin{equation}
\mathcal{T}_{n}^{<}\overrightarrow{f_{i,j,l}}=C\left[\begin{array}{rrrcrr}
0\\
 & \ddots\\
 &  & 0\\
 &  &  & T_{n}\\
 &  &  &  & \ddots\\
 &  &  &  &  & 2T_{n}\end{array}\right]\left[\begin{array}{c}
f_{i,1,l}\\
\vdots\\
f_{i,\frac{N_{k_{z}}}{2},l}\\
f_{i,\frac{N_{k_{z}}}{2}+1,l}\\
\vdots\\
f_{i,N_{k_{z}},l}\end{array}\right]_{l},\end{equation}
$\mathcal{T}_{n}^{\lessgtr}$ being a diagonal square matrix of size
$N_{k}$, and $C=\frac{\hbar\bigtriangleup k_{z}}{4m^{*}\bigtriangleup z}$.
The values of $T_{0}^{\lessgtr}=\pm3$, $T_{1}^{\lessgtr}=\mp4$,
$T_{2}^{\lessgtr}=\pm1$ are defined by their position in the complete
matrix, as stated by the second order differencing scheme and the
boundaries. By denoting $[f]_{i}$ as a vector of length $N_{k}$
holding all the momentum (j) values of $f_{ij}$ for a given i, the
entire $\mathbf{T}\cdot\overrightarrow{f}$ term is written as\begin{gather}
\left[\mathbf{T}\cdot\overrightarrow{f}\right]_{l}=C_{j}\left(\left[\begin{array}{lllll}
\mathbf{\mathcal{T}}_{0}^{<} & \mathbf{\mathcal{T}}_{1}^{<} & \mathcal{T}_{2}^{<}\\
 & \ddots & \ddots & \ddots\\
 &  & \mathbf{\mathcal{T}}_{0}^{<} & \mathbf{\mathcal{T}}_{1}^{<} & \mathcal{T}_{2}^{<}\\
 &  &  & \mathbf{\mathcal{T}}_{0}^{<} & \mathbf{\mathcal{T}}_{1}^{<}\\
 &  &  &  & \mathbf{\mathcal{T}}_{0}^{<}\end{array}\right]-\right.\nonumber \\
\left.\left[\begin{array}{lllll}
\mathcal{T}_{0}^{>}\\
\mathbf{\mathcal{T}}_{1}^{>} & \mathcal{T}_{0}^{>}\\
\mathbf{\mathcal{T}}_{2}^{>} & \mathbf{\mathcal{T}}_{1}^{>} & \mathcal{T}_{0}^{>}\\
 & \ddots & \ddots & \ddots\\
 &  & \mathbf{\mathcal{T}}_{2}^{>} & \mathbf{\mathcal{T}}_{1}^{>} & \mathcal{T}_{0}^{>}\end{array}\right]\right)\left[\begin{array}{c}
[f]_{1,l}\\
\vdots\\
\vdots\\
\vdots\\
\left[f\right]_{1,N_{k_{r}}}\end{array}\right]_{l},\label{eq:1D_T_Matrix}\end{gather}
where $C_{j}=\left(2j-N_{k_{z}}-1\right)C$, and $\mathbf{T}_{l}$
is a block tri-diagonal square matrix of rank $N_{z}N_{k_{z}}$.

Concerning the radial terms, no matrix operations are needed. Each
ring receives the Wigner function distribution from it's nearest inner
and outer neighbor and the first order differencing scheme (described
above) is used.

\subsection{Potential Matrix}

\begin{flushleft}Now, when $[f(i,j)]_{l'}$ is expressed in vector
form (for a given $r$ and $k_{r}$) as $\left[f_{1,1}f_{1,2}f_{1,3}\cdots f_{1,N_{k_{z}}}f_{2,1}\cdots f_{i,j-1}f_{i,j}f_{i,j+1}\cdots f_{N_{z},N_{k_{z}}}\right]_{l'}^{T}$,
we can write equation \ref{eq:2D_Potential_Discrete} for a given
$j$ as\end{flushleft}

\noindent \begin{widetext}

{\footnotesize \[
\left[\mathbf{V}(i,l')\cdot\overrightarrow{f_{i,j}}\right]_{n,l'}=C\left[\begin{array}{ccccc}
V_{n,l}(i,1-1,l') & V_{n,l}(i,1-2,l') & \cdots & V_{n,l}(i,1-[N_{k_{z}}-1],l') & V_{n,l}(i,1-N_{k_{z}},l')\\
V_{n,l}(i,2-1,l') & V_{n,l}(i,2-2,l') & \cdots & V_{n,l}(i,2-[N_{k_{z}}-1],l') & V_{n,l}(i,2-N_{k_{z}},l')\\
\vdots & \vdots & \ddots & \vdots & \vdots\\
V_{n,l}(N_{k_{z}}-1-1,l') & V_{n,l}(i,N_{k_{z}}-1-2,l') & \cdots & V_{n,l}(i,N_{k}-1-[N_{k_{z}}-1],l') & V_{n,l}(i,N_{k_{z}}-1-N_{k_{z}},l')\\
V_{n,l}(i,N_{k_{z}}-1,l') & V_{n,l}(i,N_{k_{z}}-2,l') & \cdots & V_{n,l}(i,N_{k}-[N_{k_{z}}-1],l') & V_{n,l}(i,N_{k_{z}}-N_{k_{z}}],l')\end{array}\right]\left[\begin{array}{c}
f_{i,1}\\
f_{i,2}\\
\vdots\\
f_{i,N_{k_{z}-1}}\\
f_{i,N_{k_{z}}}\end{array}\right]_{l'},\]
}where $V_{n,l}(i,j-j',l')=\sum_{n'}\mathcal{J}(n',l,l')\sum\limits _{i'}\sigma(i',j-j')\mathcal{V}(i,i',n,n')$
and $C=\frac{\pi^{2}}{\hbar N_{k_{r}}^{2}N_{k_{z}}}$. Since $V_{n,i}(i,j,l')\propto\sin j$,
$V_{n,l}(i,-j,l')=-V_{n,l}(i,j,l')$ and $V_{n,l}(i,0,l')=0$, the
above becomes{\footnotesize \begin{equation}
\left[\mathbf{V}(i,l')\cdot\overrightarrow{f_{i,j}}\right]_{n,l'}=C\left[\begin{array}{ccccc}
0 & -V_{n,l}(i,1,l') & \cdots & -V_{n,l}(i,N_{k_{z}}-2,l') & -V_{n,l}(i,N_{k_{z}}-1,l')\\
V_{n,l}(i,1,l') & 0 & \cdots & -V_{n,l}(i,N_{k_{z}}-3,l') & -V_{n,l}(i,N_{k_{z}}-2,l')\\
\vdots & \vdots & \ddots & \vdots & \vdots\\
V_{n,l}(i,N_{k_{z}}-2,l') & V_{n,l}(i,N_{k_{z}}-3,l') & \cdots & 0 & -V_{n,l}(i,1,l')\\
V_{n,l}(i,N_{k_{z}}-1,l') & V_{n,l}(i,N_{k_{z}}-2,l') & \cdots & V_{n,l}(i,1,l') & 0\end{array}\right]\left[\begin{array}{c}
f_{i,1}\\
f_{i,2}\\
\vdots\\
f_{i,N_{k_{z}-1}}\\
f_{i,N_{k_{z}}}\end{array}\right]_{l'}\end{equation}
}Note that $\mathbf{V}(i,l')$ is a $N_{k_{z}}\times N_{k_{z}}$ anti-symmetric
matrix.

\end{widetext}
By denoting $[f]_{i,l}$ as a vector of
length $N_{k_{z}}N_{k_{r}}$ holding all the momentum (j) values of
$f_{ijl}$ for a given $i,l$, the entire $\mathbf{U}\cdot\overrightarrow{f}$
term is written as\\
\begin{equation}
\left[\mathbf{U}\cdot\overrightarrow{f}\right]_{n}=\left[\begin{array}{rrrc}
\mathbf{V}(1,1) & \mathbf{V}(1,2) & \cdots & \mathbf{V}(1,N_{k_{r}})\\
\mathbf{V}(2,1) & \mathbf{V}(2,2) & \cdots & \mathbf{V}(2,N_{k_{r}})\\
\vdots & \vdots & \ddots & \vdots\\
\mathbf{V}(N_{z},1) & \mathbf{V}(N_{z},2) & \cdots & \mathbf{V}(N_{z},N_{k_{r}})\end{array}\right]\left[\begin{array}{c}
[f]_{1,1}\\
\left[f\right]_{1,2}\\
\vdots\\
\left[f\right]_{1,N_{k_{r}}}\end{array}\right],\label{eq:2D_Potential_Matrix}\end{equation}
so that $\mathbf{U}$ is a square matrix of rank $N_{z}N_{k_{z}}N_{k_{r}}$.

\subsection{Interaction Matrix}

Whereas we previously wrote the discrete interaction term in equation
\ref{eq:2D_Interaction_Discrete}, we rewrite it as\begin{gather}
\left[\mathbf{S}\cdot\overrightarrow{f}\right](n,i,j,l)\equiv\frac{\beta(n,i,j,l)}{\tau}\rho(n,i)-\frac{1}{\tau}f(n,i,j,l)\nonumber \\
\rho(n,i)\equiv\sum_{j'=1}^{N_{k_{z}}}\sum_{l'=1}^{N_{k_{r}}}\left|(2l'-N_{k_{r}}-1)\right|f(n,i,j',l')\nonumber \\
\beta(n,i,j,k)=\frac{f_{0}(n,i,j,l)}{\sum_{j'=1}^{N_{k_{z}}}\sum_{l'=1}^{N_{k_{r}}}\left|(2l'-N_{k_{r}}-1)\right|f_{0}(n,i,j',l')},\end{gather}
where $\rho(n,i)$ is the density of the previous time step.

By writing$f(i,j)$ as a vector, $\left[f_{1,1}f_{1,2}f_{1,3}\cdots f_{1,N_{k}}f_{2,1}\cdots f_{i,j-1}f_{i,j}f_{i,j+1}\cdots f_{N_{x},N_{k}}\right]^{T}$
and by denoting $[f]_{i}$ as a vector of length $N_{k}$ holding
all the momentum (j) values of $f_{ij}$ for a given i, the entire
$\mathbf{S}\cdot\overrightarrow{f}$ term is written as\begin{gather}
\left[\overrightarrow{\Sigma}\right]_{n,l}-\left[\mathbf{S}\cdot\overrightarrow{f}\right]_{n,l}=\frac{1}{\tau}\left[\begin{array}{c}
[\rho]_{1}[\beta]_{1}\\
{}[\rho]_{2}[\beta]_{2}\\
\vdots\\
{}[\rho]_{N_{z}}[\beta]_{N_{z}}\end{array}\right]-\nonumber \\
\left[\begin{array}{rrrc}
S(1) & 0 & \cdots & 0\\
0 & S(2) & \cdots & 0\\
\vdots & \vdots & \ddots & \vdots\\
0 & 0 & \cdots & S(N_{z})\end{array}\right]\left[\begin{array}{c}
[f]_{1}\\
{}[f]_{2}\\
\vdots\\
{}[f]_{N_{z}}\end{array}\right],\label{eq:2D_Interaction_Matrix}\end{gather}
where $\mathbf{S}$ is a square block diagonal matrix of rank $N_{Z}N_{k_{z}}$
and $\overrightarrow{\Sigma}$ is a $N_{z}N_{k_{z}}$ vector.

\subsection{Boundary Conditions}

By denoting $[f]_{i}$ as a vector of length $N_{k}$ holding all
the momentum (j) values of the Fermi distribution $f_{Fermi}(j)$
for a given i, the longitudinal boundary equations (\ref{eq:BC-begin}
to \ref{eq:BC-end}) can be written as\begin{equation}
\overrightarrow{\mathbf{B}}=C_{j}\left[\begin{array}{c}
B_{1}^{>}[f]_{1}\\
B_{2}^{>}[f]_{2}\\
\vdots\\
B_{2}^{<}\left[f\right]_{N_{z}-1}\\
B_{1}^{<}\left[f\right]_{N_{z}}\end{array}\right],\end{equation}
$B_{n}^{\lessgtr}[f]_{i}$ being a vector of size $N_{k_{z}}$, $\overrightarrow{\mathbf{B}}$
a vector of size $N_{z}N_{k_{z}}$, and $C_{j}=\frac{\hbar\bigtriangleup k_{z}}{4m^{*}\bigtriangleup z}\left(2j-N_{k_{z}}-1\right)$.
The values of $B_{1}^{\lessgtr}=\pm2$, $B_{2}^{\lessgtr}=\mp1$ are
defined by equations \ref{eq:BC-begin} to \ref{eq:BC-end}, as stated
by the second order differencing scheme at the boundaries.

\section{Methods of Solution\label{sec:Methods-of-Solution}}

\subsection{Parallelization}

So far, we have dealt almost exclusively with the 1D space, 2D momentum
transport problem (1x+2k). As stated above, we are working under the
assumption that \emph{longitudinal transport is more dominant than
radial transport}. This allows the total transport to be calculated
in two steps: (1) transport the particles in the longitudinal direction
in each shell separately (1x+2k), then (2) each shell exchanges particles
with its nearest neighbor. This latter step is where we employ parallel
processing techniques. The 1x+2k problem is performed on P processors,
where P is the number of cylindrical shells into which we have divided
up the RTD. Once each shell has advanced a given amount of time, then
communication between shells (radial drift) can commence. As described
in Section \ref{sub:Drift-&-Boundary}, a central differencing scheme
(CDS) is used and shown in equations \ref{eq:radial_drift_in}-\ref{eq:radial_drift_out}.
The boundaries consist of the material external to the shell and the
innermost shell. As per equations \ref{eq:BC-inner-1}-\ref{eq:BC-outer-in}
and explained in section \ref{sub:Drift-&-Boundary}, the exterior
boundary defines the device. For example, if the device is a mesa
RTD, and there is nothing but vacuum outside the shell, one should
choose a boundary shell that injects into the outermost shell the
same particles that the outermost shell ejected (keeping the momenta
the same). If the device is a slab with a circular contact for an
emitter, then the material outside the cylinder is in equilibrium.
The boundary shell would be chosen to reflect this.

\subsection{Potential Transform\label{sub:Potential-Transform}}

As explained is section \ref{sec:2D-Matrix-Setup} and seen in equation
\ref{eq:2D_Potential_Matrix}, the Wigner integral (potential term)
is the only term that can not be made diagonal in $k_{r}$, leading
to a full matrix that is too big to store in memory. If one uses direct
integration methods (\cite{ROCK4} ), the number of terms becomes
large enough to make the problem intractable. We now describe a method
of using the Fourier transform property of the Wigner integral to
eliminate the $k_{r}$ dependence inherent in equation \ref{eq:F}.

We need to have either the matrix $\Omega=\mathbf{T}_{z}+\mathbf{U}+\mathbf{S}$
be able to be stored in the RAM of present day computers (for the
matrix methods), or limit the number of simultaneous equations to
solve ( the direct integration methods). The idea behind our method
of solution of the 2D transport equation, \begin{equation}
\frac{d\mathbf{f}}{dt}=\Omega\mathbf{f}-\mathbf{B},\label{eq: WFmatrixE}\end{equation}
is that if the matrix $\Omega$ is block diagonal in $k_{r}$ then
one can progress through all the values of $k_{r}$ solving the matrix
equation at fixed values of $k_{r}$ each time, effectively reducing
the simulation to a series of 1D problems. Unfortunately, the equations
for some of the matrix operators are not block diagonal in $k_{r}$
in their present form. Some manipulation will be needed to obtain
block diagonal terms. The drift term is already independent of $k_{r}$,
and the scattering term is coupled to $k_{r}$ via off-diagonal elements
due to the integral $\int\left|k'_{r}\right|dk'_{r}\int dk'_{z}\, f(r,z,k'_{z},k'_{r})$.
This integral is just the 1D density in the shell $\rho_{r}(z)$,
which can be calculated at the beginning of the time step. This approximation
makes the scattering term into the desired diagonal matrix without
losing too much detail. The potential term, however, is not so simple.
In equation (\ref{eq:Uf-continuous}), the term $\mathcal{F}(r,z,\rho,\zeta)$,
as defined in equation (\ref{eq:F}) makes the matrix operator $\mathbf{U}$
a full matrix. We will now outline a method to circumvent this problem
below.

In order to solve the transport equation, we use an implicit method
(first proposed in \cite{Buot-Jensen-PRB-1990}) by rewriting equation
\ref{eq: WFmatrixE} as (dropping the $z$ index from the potential
matrix)\begin{equation}
\frac{\overline{\mathbf{f}}-\mathbf{f}}{\Delta t}=\left(\mathbf{T}+\mathbf{U}+\mathbf{S}\right)\frac{\overline{\mathbf{f}}+\mathbf{f}}{2}-\mathbf{B},\end{equation}
where $\overline{\mathbf{f}}$ means the new (next) value of $\mathbf{f}$
in time. Rewriting it as\begin{equation}
\left[1-\frac{\Delta t}{2}\left(\mathbf{T}+\mathbf{U}+\mathbf{S}\right)\right]\cdot\left(\overline{\mathbf{f}}+\mathbf{f}\right)=2\mathbf{f}+\mathbf{B}\Delta t,\end{equation}
and making the approximation (accepting error in terms of order $\Delta t^{2}$)\begin{equation}
1-\frac{\Delta t}{2}\left(\mathbf{T}+\mathbf{U}+\mathbf{S}\right)\simeq\left[1-\frac{\Delta t}{2}\left(\mathbf{T}+\mathbf{S}\right)\right]\left[1-\frac{\Delta t}{2}\mathbf{U}\right]\label{eq:approx}\end{equation}
 allows us to write that \begin{equation}
\left[1-\frac{\Delta t}{2}\left(\mathbf{T}+\mathbf{S}\right)\right]\left[1-\frac{\Delta t}{2}\mathbf{U}\right]\cdot\left(\overline{\mathbf{f}}+\mathbf{f}\right)=2\mathbf{f}+\mathbf{B}\Delta t.\label{eq:split_eqn}\end{equation}
For convenience, we will rewrite this as \begin{equation}
\Omega\mathbf{f'}=\Omega_{0}\Omega_{U}\mathbf{f}'=\mathbf{b},\end{equation}
where we have defined\begin{eqnarray}
\mathbf{b} & = & 2\left(\mathbf{f}+\mathbf{B}\tau\right)\\
\tau & = & \frac{\Delta t}{2}\\
\Omega_{U} & = & \left(1-\tau\mathbf{U}\right)\\
\Omega_{0} & = & \left(1-\tau\left[\mathbf{T}+\mathbf{S}\right]\right)\\
\mathbf{f}' & = & \left(\overline{\mathbf{f}}+\mathbf{f}\right),\end{eqnarray}

The solution to this equation involves solving two matrix equations.

\begin{enumerate}
\item Solve $\Omega_{0}\Gamma=\mathbf{b}$ for $\Gamma$ (quick)
\item Solve $\Omega_{U}\mathbf{f}'=\Gamma$ for $\mathbf{f}'$ (impractical)
\end{enumerate}
\begin{flushleft}The matrix $\Omega_{U}$ is still too big to store
in memory and, consequently, solving $\Omega_{U}\mathbf{f}'=\Gamma$
for $\mathbf{f}$ is not practical for modern computers. By making
the approximation that led to equation (\ref{eq:split_eqn}), we are
able to solve $\Omega_{U}\mathbf{f}'=\Gamma$ separately, allowing
us to reformulate it in a form more suitable for computation. \end{flushleft}

\begin{flushleft}Recall that the Wigner integral, $\mathbf{U\cdot f}$
, was derived by taking the Fourier transforms of the Greens function,
$G^{<}$\cite[sec 6.4]{K&B}. From the last term in equation \ref{eq:WFE-2D_ptI},
can be written out in full as \begin{gather}
\int dz'e^{-\textrm{i}2k_{z}z'}\int\left|r'\right|dr'J_{0}(2k_{r}r')\mathcal{V}(z,z',r,r')\times\nonumber \\
\int dk'_{z}e^{+\textrm{i}2k'_{z}z'}\int\left|k'_{r}\right|dk'_{r}J_{0}(2k'_{r}r')f(r,z,k'_{z},k_{r}')\nonumber \\
=\mathbf{F}\mathcal{V}\mathbf{F}^{-1}\mathbf{f},\end{gather}
{\footnotesize \begin{equation}
\mathcal{V}(z,z',r,r')=\int_{0}^{2\pi}d\theta\left\{ V\left(z+z',r+r'\cos\theta\right)-V\left(z-z',r-z'\cos\theta\right)\right\} .\end{equation}
}The Fourier Bessel transform,$\mathbf{F}$, and its inverse, $\mathbf{F}^{-1}$,
are given by {\small \begin{gather}
\mathbf{F}\equiv\mathbf{F}(k_{r},k_{z};r',z')=\int dz'e^{-\textrm{i}2k_{z}z'}\int dr'\left|r'\right|J_{o}\left(2k_{r}r'\right)\\
\mathbf{F}^{-1}\equiv\mathbf{F}^{-1}(k_{r},k_{z};r',z')=\frac{1}{(2\pi)^{2}}\int dk_{z}e^{+\textrm{i}2k'_{z}z'}\int dk_{r}\left|k_{r}\right|J_{0}(2k_{r}r').\end{gather}
}When we define\begin{equation}
g(\mathbf{x},\mathbf{y})=\frac{1}{(2\pi)^{3}}\int d^{3}\mathbf{k}\, e^{2i\mathbf{k}\cdot\mathbf{y}}f(\mathbf{x},\mathbf{k})=\mathbf{F}^{-1}\mathbf{f}.\end{equation}
we can rewrite $\Omega_{U}\mathbf{f}'=\Gamma$ as\end{flushleft}

\begin{eqnarray}
\left(1-\tau\mathbf{F}\mathcal{V}\mathbf{F}^{-1}\right)\left(\mathbf{Fg}'\right) & = & \Gamma,\end{eqnarray}
 $\mathbf{g}$ being the Fourier Transform of $\mathbf{f}$ and $\mathbf{g}'=\overline{\mathbf{g}}+\mathbf{g}$.
Some manipulation gives\begin{eqnarray}
\Omega'_{U}\mathbf{g}' & = & \gamma.\end{eqnarray}
where\begin{eqnarray}
\Omega'_{U} & = & \left(1-\tau\mathcal{V}\right)\\
\gamma & = & \mathbf{F}^{-1}\Gamma.\end{eqnarray}
The new procedure is to solve for the new equation, $\Omega'_{U}\mathbf{g}'=\gamma$.

The term $\mathcal{F}(r,z,\rho,\zeta)$, as defined in equation (\ref{eq:F}),
can now be written as \begin{eqnarray}
\mathcal{F}(r,z,\rho,\zeta) & = & \int\left|k'_{r}\right|dk'_{r}dk'_{z}\int_{0}^{2\pi}d\chi'_{\phi}e^{+2\textrm{i}(k'_{z}\zeta+k'_{r}\cos\chi'_{\phi}\rho)}\times\nonumber \\
 &  & f(r,z,k'_{z},k'_{r},\chi'_{\phi})\nonumber \\
 & = & \int d^{3}\mathbf{k'}e^{2\textrm{i}\mathbf{k}\cdot\mathbf{y}}f(r,z,\mathbf{k})\nonumber \\
 & = & (2\pi)^{3}g(r,z,\rho,\zeta,\theta).\end{eqnarray}
This lets us define $\Omega'_{U}=\left[1-\frac{\Delta t}{2}\mathbf{U}'\right]$
in terms of\begin{gather}
\mathbf{U'g}=\frac{1}{2\pi^{2}\hbar}\int d\rho d\zeta\left|\rho\right|\sin\left(2k_{z}\zeta\right)J_{0}\left(2k_{r}\rho\right)\times\nonumber \\
\mathcal{V}(r,z,\rho,\zeta)(2\pi)^{3}g(r,z,\rho,\zeta,\theta),\end{gather}
or, in discrete form, as (completing equation (\ref{eq:2D_Potential_Discrete})
) {\scriptsize }\begin{gather}
\left[\mathbf{U'g}\right](n,i,j,k)=+\frac{16\pi}{\hbar}\Delta r^{2}\Delta z\sum_{n',i'}\mathcal{U}(n,i,n',i',k,j)\times\nonumber \\
g(n,i,n',i',m').\end{gather}
The WDF is recovered by \begin{eqnarray}
f(r,z,k_{r},k_{z}) & = & \int_{0}^{2\pi}\left|k_{r}\right|d\chi_{\phi}f(r,z,k_{r},k_{z},\chi_{\phi})\nonumber \\
 & = & \int_{0}^{2\pi}\left|k_{r}\right|d\chi_{\phi}\int d^{3}\mathbf{y}\, e^{2i\mathbf{k}\cdot\mathbf{y}}g(r,z,\mathbf{y})\nonumber \\
 & = & \left|k_{r}\right|\int d\rho\left|\rho\right|J_{0}(2k_{r}\rho)\int d\zeta e^{-2ik_{z}\zeta}\times\nonumber \\
 &  & g(r,z,\rho,\zeta).\end{eqnarray}
The complete process to solve the WFE equation is given in the following
steps

\begin{enumerate}
\item Solve $\Omega_{0}\Gamma=\mathbf{b}$ for $\Gamma$
\item Define $\gamma=\mathbf{F}^{-1}\Gamma.$
\item Solve $\Omega'_{U}\mathbf{g}'=\gamma$ for $\mathbf{g}'$
\item Recover $\mathbf{f}'$ from $\mathbf{f}'=\mathbf{Fg}'$
\end{enumerate}

\subsection{Aim and shoot}

We find that using direct integration for a system of linear equations
grew slower and less stable as the number of simultaneous equations
grew. For example, a 1fs time step involved hundreds of iterations
were involved, each one taking about 90 seconds. When using an implicit
matrix method, as is standard in the 1D codes, the matrices were too
big to store. Because we have now, using the approximation in equation
\ref{eq:approx}, separated the operator $\Omega$ into the product
of a drift/scattering operator, $\Omega_{0}$, and a potential operator,
$\Omega_{U}$, a combination of these two methods can be used. As
we shown, by itself, $\Omega_{0}$ can be block diagonal and sparse
enough to be easily solved by either of the two methods. Also, by
itself, $\Omega_{U}$ can be rewritten in a form that also can be
easily solved by either of the two methods. 

The following method allows us to take advantage of this split is
a way that is analogous to the accelerated convergence method used
to obtain steady state solutions in the 1D problem \cite{Buot-Jensen-PRB-1990}.
We have dubbed this a {}``aim and shoot'' approach. The static potential
term is calculated on a time scale of $\delta t=0.01$ fs for only
one step, then held constant while the system evolves for $\Delta t=1$fs.
This is the aim part. The shoot part involves the drift and scattering
matrices being solved implicitly by matrix inversion, with the potential
term static. This is fine for a steady-state solution, but it is still
uncertain if this method will give the proper transient behavior of
the system.

\section{Implementation and simulation results}

In the preceding sections a computational method of solving for both
the time dependent and steady state two dimensional Wigner function
transport equation was presented. The 2D equations and computational
method were derived for the case of longitudinal transport through
a cylinder while taking account of the effects of radial momentum
in addition to the longitudinal momentum. As previously stated, the
numerical solution is broken into two parts: (1) transport in the
longitudinal direction then (2) transport in the radial direction.
The cylinder is divided in to a number of concentric cylindrical shells
in which the longitudinal transport takes place as in the 1D problem,
but with the inclusion of radial momentum. The radial transport involves
a simple exchange of particles (dictated by the newly calculated radial
momentum). Sine this step is computationally trivial, this work was
concerned with the former step: A 1D space and 2D momentum transport
problem (1x+2k). Below we present some proof-of-principle simulation
results obtained using the methods developed in section \ref{sec:Methods-of-Solution}.
From these results, the future usefulness of each of these methods,
in light of current computing trends, will be discussed.

This simulations were carried mostly on Linux workstations (2GHz Pentium4,
1.7GHz Pentium4s and 1.2GHz AMD AthlonMPs). For phase space, our solution
method is most easily formulated when $N_{z}=N_{k_{z}}$ and $N_{r}=N_{k_{r}}$
since the Fourier transforms between momentum space and displacement
space must be on the same lattice. Simulations were performed on longitudinal
phase space grid sizes of $N_{z}=N_{k_{z}}=96$ and $N_{z}=N_{k_{z}}=48$
with radial phase space grid sizes of $N_{r}=N_{k_{r}}=2,4,8,16,20$.
In all of the simulations presented, the device is constructed of
bulk n-doped GaAs with a RTS of undoped $GaAs/Al_{0.3}Ga_{0.7}As$
with a barrier potential is 0.3 eV. The device temperature is 77K,
the electron effective mass is 0.0667$m_{0}$ and The donor density
is $2\times10^{18}cm^{-3}$. The cylinder has dimensions of $1000\textrm{Å}$
in both length (z) and diameter (r). For most of the simulations the
active RTS region is approximately $180\textrm{Å}$. The well, spacer
and barrier lengths will be given for each simulation.

\begin{figure}
\noindent \begin{center}\begin{tabular}{cc}
\includegraphics[%
  width=0.50\columnwidth,
  keepaspectratio]{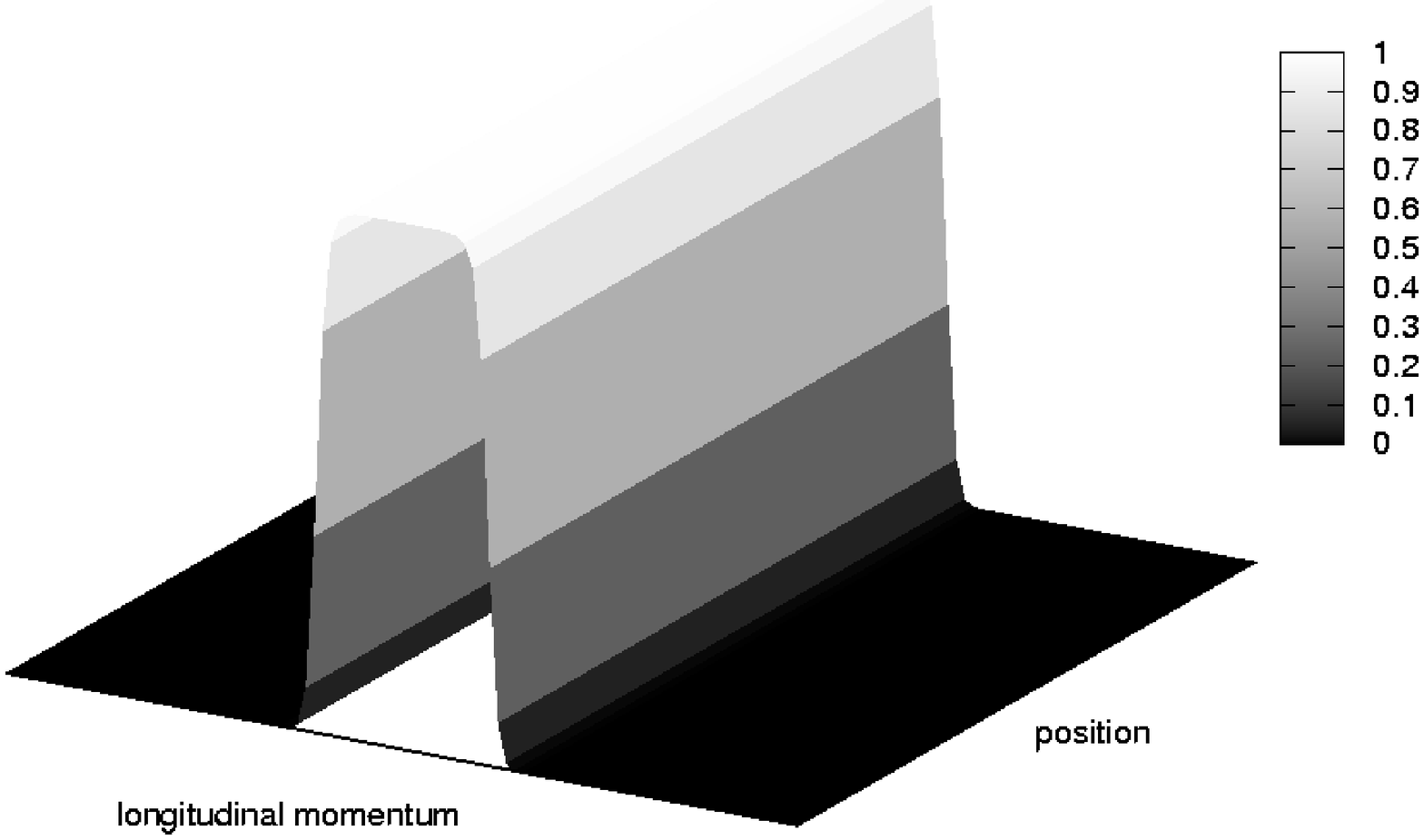}&
\includegraphics[%
  width=0.50\columnwidth,
  keepaspectratio]{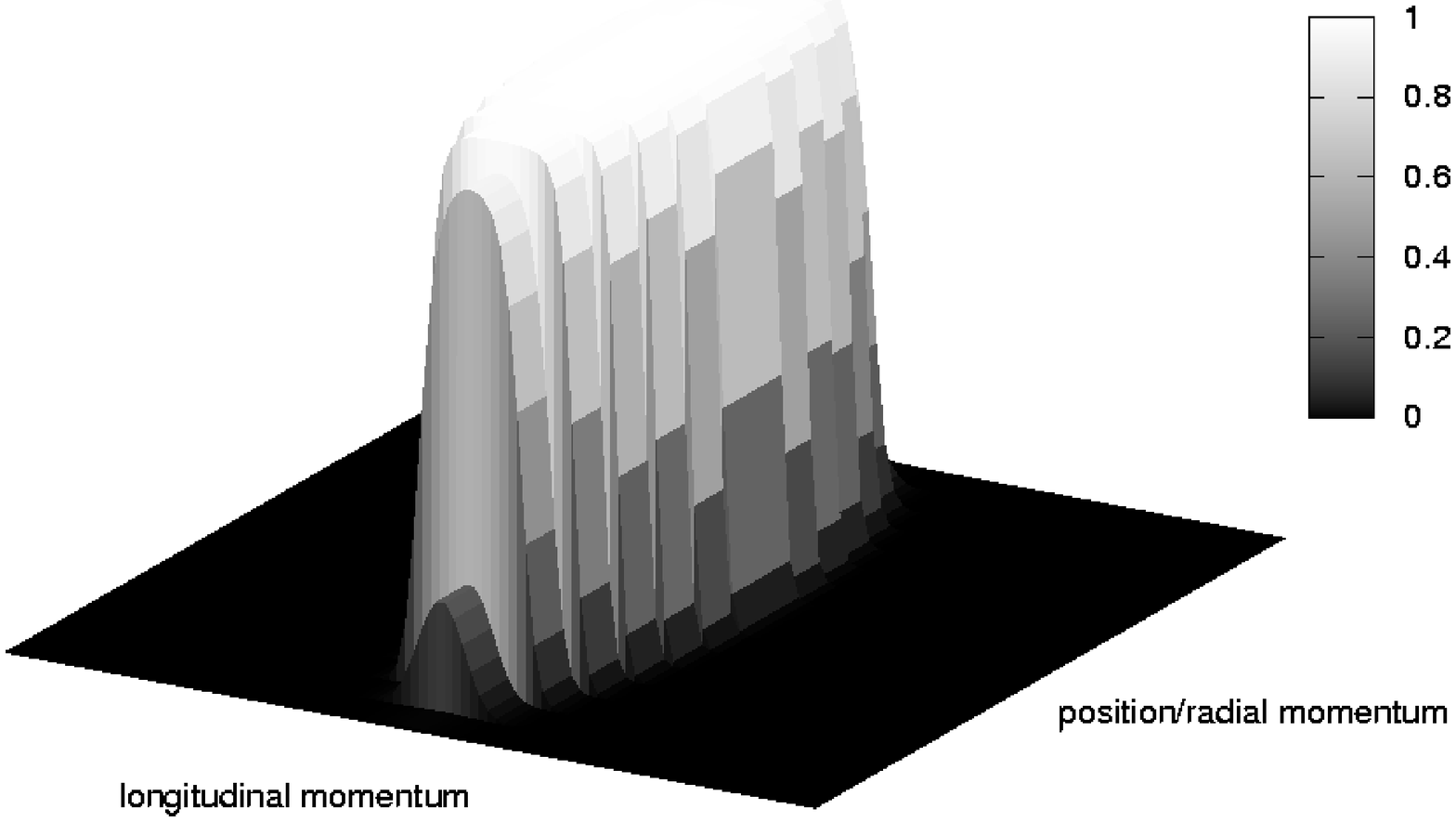}\tabularnewline
{\small (a)}&
{\small (b)}\tabularnewline
\end{tabular}\end{center}

\caption{{\small \label{cap:2D-Fermi-WDF}Initial WDF as Boundary Value. (a)
Longitudinal phase space for one $k_{r}$ slice. (b) All $k_{r}$
slices put together.}}
\end{figure}

\begin{figure}
\begin{center}\includegraphics[%
  width=1.0\columnwidth,
  keepaspectratio]{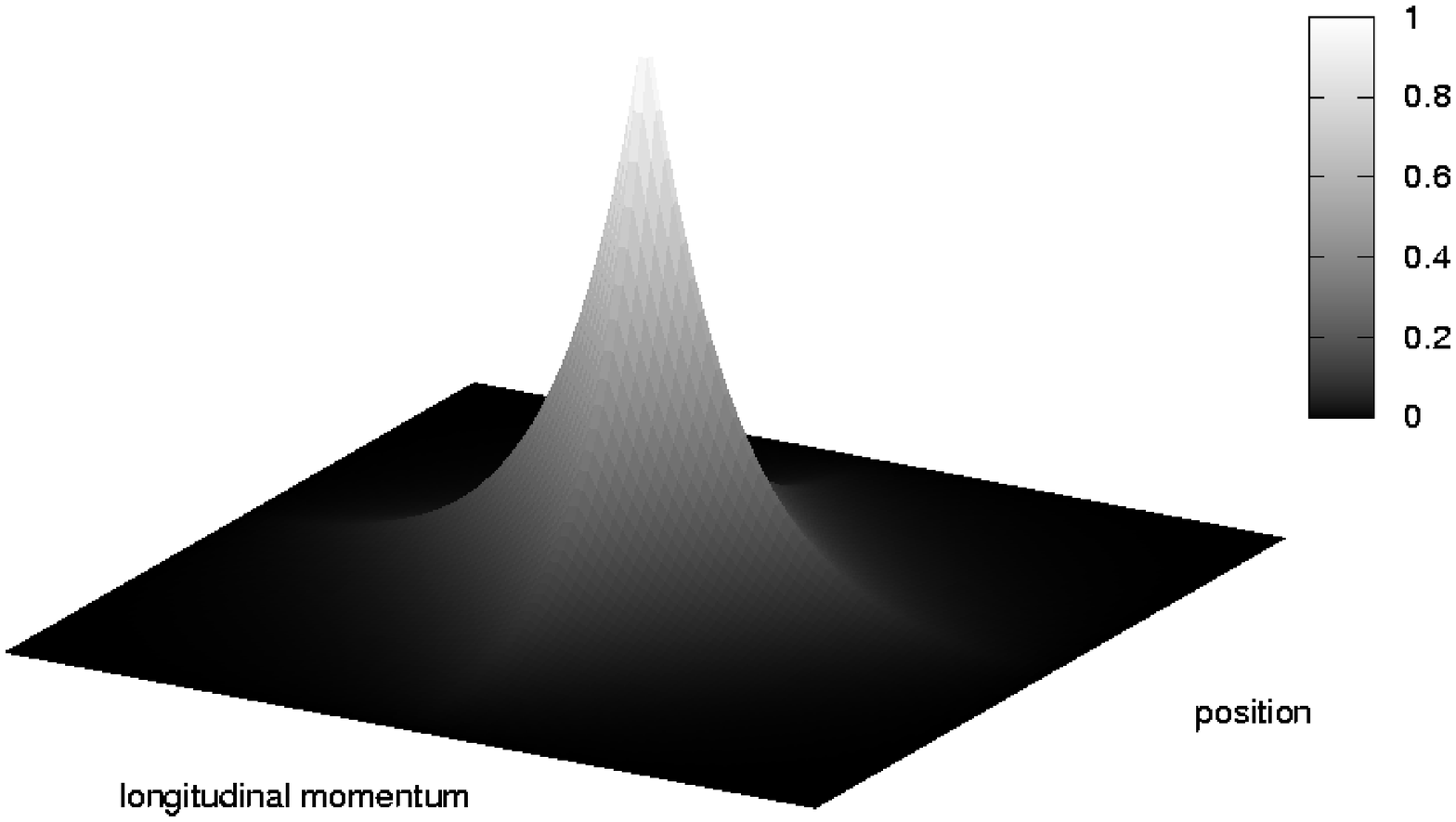}\end{center}

\caption{\label{cap:2D-Special-WDF}Initial WDF as a fanciful test distribution.}
\end{figure}

Each numerical experiment has been carried out in the following way.
First, the cylinder is populated with electrons according to one of
three specific WDF: (a) $f(z,k_{Z},k_{r})=0$, corresponding to no
excess electrons in the cylinder, (b) $f(z,k_{Z},k_{r})=f_{Fermi}$,
corresponding to the same distribution of the metallic leads (figure
\ref{cap:2D-Fermi-WDF}), and (c) a fanciful distribution corresponding
to electrons mostly at the center of phase space (figure \ref{cap:2D-Special-WDF}).
To be exact, an error was made in preparing for case (b). We meant
to set the WDF each $k_{r}$ slice to the Fermi distribution of the
boundary of that specific $k_{r}$ slice. This way the integral of
the total WDF would be unity, as expected. Instead, we accidentally
normalized the WDF of each $k_{r}$slice such that the integral of
the WDF in each slice in unity. We have kept this error here for the
reason that the simulations illustrates that the system will adjust
itself and still tend toward the expected result. %
 Next, at zero bias, the system is allowed to evolve with scattering
turned off. The system is allowed to evolve for a suitable time, until
it settles into a steady state, and then scattering is turned on.
During this time, the previous step's value of the WDF is used as
the {}``equilibrium value'' needed in the relaxation time approximation
of the present step. Once again, the system is allowed to evolve for
a suitable time, and at this time, the current WDF is set to be the
{}``equilibrium value'' for the rest of the evolution, which continues
until the system settles into a steady state. By observing this time
evolution of each of the three cases, we can determine how well a
given method behaves as well as gather timing and other information.

One important item of note is that the examples below are performed
with $N_{z}=N_{k_{z}}=96$ and $N_{r}=N_{k_{r}}=2$. In effect we
are including only one positive and one negative radial momentum value.
While this is fine for testing purposes where we are basically reducing
the simulation to a 1D problem, for a real 2D problem $N_{k_{r}}$
and $N_{r}$ should be large enough (\textasciitilde{}16) to encompass
a phase space greater than the radial Fermi momentum of the material
outside the cylinder. The computational issue is that while $N_{r}=N_{k_{r}}=2$
can be solved in under 20min for a 2000fs run at time steps of 1fs,
the addition of more radial points increases the time dramatically
(this will be mentioned below). As a result, the phase space plots
given are showing only one value of the radial momentum (the positive
momentum) since they are symmetric in this case.

In order to follow the time evolution of this system, we have employed
two different types of integration methods: Direct integration (explicit)
and Matrix solvers (implicit). Direct integration is done using a
pre-packaged integrator called ROCK4\cite{ROCK4}, which is a $4^{th}$
order Runge-Kutta like integrator for a system of equations. With
ROCK4, there is no need to compute and store the right hand side of
the discretized equation as a general band matrix, and consequently,
no need for a matrix inversion of the time evolution operator which
have been used in previous simulations. Instead, all that is needed
is to calculate the right hand side on the fly. Implicit integration
is done by using LAPACK\cite{LAPACK} to solve the matrices.

\subsection{Direct Integration Methods}

Not much will be said for the explicit method since, on the same computer
as the runs below, after 36 minutes the method progressed only to
a time of 14fs. When the number of radial grid points is increased
from 2 up to 8, the number of equations increases 256 times the original
number. This fact renders ROCK4 useless for any future 2D simulations.
Recently, we have been introduced\cite{Kelley} to implicit direct
integration methods (BDF/Adams) and Newton solvers that, so far, outperform
the ROCK4 method for the 1D simulations. We have yet to include this
method in our 2D simulations. The next phase of the ongoing research
is to see how such methods compare to what we will present below.

\subsection{Matrix Split\label{sub:Matrix-Split-numerics}}

The so-called matrix split refers to the method of section \ref{sub:Potential-Transform},
which is when take the WFE, apply the approximation of equation \ref{eq:approx}
and split the matrix $\Omega$ into a product of a drift/scattering
matrix and a potential matrix (equation \ref{eq:split_eqn}). This
can be written as \begin{equation}
\Omega\mathbf{f'}=\Omega_{0}\Omega_{U}\mathbf{f}'=\mathbf{b},\end{equation}
The solution to this equation involves solving two matrix equations,
one for the drift/scattering\begin{equation}
\Omega_{0}\Gamma=\mathbf{b},\end{equation}
and one for the potential\begin{equation}
\Omega'_{U}\mathbf{g}'=\gamma.\end{equation}
In solving the potential equation, we must perform an inverse transform,
$\gamma=\mathbf{F}^{-1}\Gamma$, and then a transform, $\mathbf{f}'=\mathbf{Fg}'$.

Figures \ref{Matrix-NoU-10} through \ref{Matrix-withU-2000} illustrates
the time evolution of our device up to 2000fs. Each figure is a snapshot
in the evolution of three WDFs whose initial values are one of the
three discussed above. They will be referred to the Zero, PseudoFermi
and Central initial values. In each figure, the left column shows
the WDF plotted in 3D, while the right column shows a contour plot
of the WDF. We will examine two distinct sets of cases to illustrate
our method. The first will have the potential term set to zero, corresponding
to no coulombic interaction. The second will include the coulombic
effects. 

\begin{table}
\begin{center}\begin{tabular}{|c|c||c|c|c|c|}
\cline{2-2} \cline{3-4} \cline{5-6} 
\multicolumn{1}{c|}{}&
Potential&
\multicolumn{2}{c|}{$U\ne0$}&
\multicolumn{2}{c|}{$U=0$}\tabularnewline
\cline{2-2} \cline{3-4} \cline{5-6} 
\cline{2-2} \cline{3-3} \cline{4-4} \cline{5-5} \cline{6-6} 
\multicolumn{1}{c|}{}&
\# time steps&
$200\Delta t$&
$2000\Delta t$&
$200\Delta t$&
\multicolumn{1}{c|}{$2000\Delta t$}\tabularnewline
\cline{2-2} \cline{3-3} \cline{4-4} \cline{5-5} \cline{6-6} 
\hline 
\multicolumn{1}{|c|}{Initial}&
Zero&
0m7.428s&
0m51.376s&
1m38.178s&
\multicolumn{1}{c|}{16m37.437s}\tabularnewline
\cline{2-2} \cline{3-3} \cline{4-4} \cline{5-5} \cline{6-6} 
\multicolumn{1}{|c|}{WDF}&
PseudoFermi&
0m7.349s&
0m51.762s&
1m36.676s&
15m45.484s\tabularnewline
\cline{2-2} \cline{3-3} \cline{4-4} \cline{5-5} \cline{6-6} 
\multicolumn{1}{|c|}{Value}&
Central&
0m7.396s&
0m52.361s&
1m35.993s&
15m38.711s\tabularnewline
\hline
\end{tabular}\end{center}

\caption{\label{cap:Matrix-Split-Timings}Timings for the Matrix Split runs}
\end{table}

\subsubsection{Potential {}``Turned Off''}

Figures \ref{Matrix-NoU-10} through \ref{Matrix-NoU-2000} show the
evolution of the electron distribution in our device without any coulomb
interactions. We see that for each of the three WDFs, the system behaves
as expected. With a zero initial WDF, we see the electrons move in
from either end of the device with the high momentum carriers further
in than the low momentum carriers. As time progresses, the phase space
fills with carriers and the WDF tends towards the distribution of
the boundaries, namely, the Fermi distribution. The next case, that
of the initial WDF set to the PseudoFermi distribution, shows the
excess carriers leaving the system. Ultimately, the phase space once
again tends towards the distribution of the boundaries. In the final
case, the Central distribution, we see the carriers moving in from
the boundary, as in the first case. At the same time, the central
distribution itself relaxes, with the positive momentum carriers moving
one direction and the negative momentum carriers the other way. Eventually,
the central distribution relaxes while the boundary carriers move
in. Once again, the end result tends towards the boundary distribution.

All three of the different initial WDFs evolve towards the same, expected,
final WDF. This simply shows that the drift/scattering part of the
simulation works as expected, which is expected. What we can learn
from this exercise is the CPU time to calculate the drift and scattering
terms up to 200 and 2000 fs (Table \ref{cap:Matrix-Split-Timings}).
As stated above, this is the first of two matrix equations that must
be solved. We see from the table that this is not where most of the
CPU will spend its time. Rather, the potential term will take the
bulk of the computing time. Next, we see how the simulation behaves
when this term is turned on.

\noindent %
\begin{figure}[H]
\begin{center}DUE TO ARXIV SIZE CONSIDERATIONS THESE GRAPHS ARE PLACED
IN A SEPARATE FILE AVAILABLE FOR DOWNLOAD ALONG WITH THIS\end{center}

\caption{\label{Matrix-NoU-10}Neglecting coulomb interactions: Surface and
Contour plots of the WDF at=10fs for the Initial WDF of (a) zero,
(b) PseudoFermi, (c) Central}
\end{figure}
\begin{figure}[H]
\begin{center}DUE TO ARXIV SIZE CONSIDERATIONS THESE GRAPHS ARE PLACED
IN A SEPARATE FILE AVAILABLE FOR DOWNLOAD ALONG WITH THIS\end{center}

\caption{\label{Matrix-NoU-50}Neglecting coulomb interactions: Surface and
Contour plots of the WDF at=50fs for the Initial WDF of (a) zero,
(b) PseudoFermi, (c) Central}
\end{figure}
\begin{figure}[H]
\begin{center}DUE TO ARXIV SIZE CONSIDERATIONS THESE GRAPHS ARE PLACED
IN A SEPARATE FILE AVAILABLE FOR DOWNLOAD ALONG WITH THIS\end{center}

\caption{\label{Matrix-NoU-100}Neglecting coulomb interactions: Surface and
Contour plots of the WDF at=100fs for the Initial WDF of (a) zero,
(b) PseudoFermi, (c) Central}
\end{figure}
\begin{figure}[H]
\begin{center}DUE TO ARXIV SIZE CONSIDERATIONS THESE GRAPHS ARE PLACED
IN A SEPARATE FILE AVAILABLE FOR DOWNLOAD ALONG WITH THIS\end{center}

\caption{\label{Matrix-NoU-200}Neglecting coulomb interactions: Surface and
Contour plots of the WDF at=200fs for the Initial WDF of (a) zero,
(b) PseudoFermi, (c) Central}
\end{figure}
\begin{figure}[H]
\begin{center}DUE TO ARXIV SIZE CONSIDERATIONS THESE GRAPHS ARE PLACED
IN A SEPARATE FILE AVAILABLE FOR DOWNLOAD ALONG WITH THIS\end{center}

\caption{\label{Matrix-NoU-1000}Neglecting coulomb interactions: Surface
and Contour plots of the WDF at=1000fs for the Initial WDF of (a)
zero, (b) PseudoFermi, (c) Central}
\end{figure}
\begin{figure}[H]
\begin{center}DUE TO ARXIV SIZE CONSIDERATIONS THESE GRAPHS ARE PLACED
IN A SEPARATE FILE AVAILABLE FOR DOWNLOAD ALONG WITH THIS\end{center}

\caption{\label{Matrix-NoU-2000}Neglecting coulomb interactions: Surface
and Contour plots of the WDF at=2000fs for the Initial WDF of (a)
zero, (b) PseudoFermi, (c) Central}
\end{figure}

\subsubsection{Potential {}``Turned On''}

Figures \ref{Matrix-withU-10} through \ref{Matrix-withU-2000} show
the evolution of the electron distribution in our device including
coulomb interactions. We can compare the simulations of these three
cases with coulombic interactions to those above, where coulombic
interactions were ignored. By following the evolution of the first
case (zero initial WDF), we see how the carriers interact with the
barriers as they move towards the center of the device. We also see
(more clearly in the contour plots) how the carriers interact with
each other, spreading out slightly as the progress inwards. In figures
\ref{Matrix-withU-100} and \ref{Matrix-withU-200} we begin to see
the interaction of the reflected carriers with the incoming carriers.
This effect grows as the system evolves, which is evident in the dark/light
patterns along the momentum axis. The same effects are seen in the
other two cases, with the exception that they begin with the unlikely
distribution having carriers in the barriers. We see these carriers
begin ejected from the barrier region at high momentum at first, then
at later times these cases evolve to same result as the first case,
where we see the expected WDF.

As we increase $N_{r}=N_{k_{r}}$ from 2 to 4, a 200 fs run increases
from about 1m40s to 5min (a factor of 3). Projecting this to 2000fs,
we see the simulation would take approximately 47 min to complete.
So far, this is not unreasonable, since for a given bias point, a
2000fs run is enough to assure convergence. A 90 point IV curve (including
the reverse sweep), would take about 70 hours (about 3 days). Increase
$N_{r}=N_{k_{r}}$ to 8 and a 200fs run takes 20min ( $4\times N_{r}=4$,
or $12\times N_{r}=2$). A 2000fs run will take $3\frac{1}{3}$hours,
which means a 90 point IV curve will take 12.5 days. A beginning run
at $N_{r}=N_{k_{r}}=16$ returns a time rate of 38 sec/fs. At that
rate, a 2000fs run would take 21.1 hours and a 90 point IV curve almost
80 days.

\begin{figure}[H]
\begin{center}DUE TO ARXIV SIZE CONSIDERATIONS THESE GRAPHS ARE PLACED
IN A SEPARATE FILE AVAILABLE FOR DOWNLOAD ALONG WITH THIS\end{center}

\caption{\label{Matrix-withU-10}Including coulomb interactions: Surface and
Contour plots of the WDF at=10fs for the Initial WDF of (a) zero,
(b) PseudoFermi, (c) Central}
\end{figure}
\begin{figure}[H]
\begin{center}DUE TO ARXIV SIZE CONSIDERATIONS THESE GRAPHS ARE PLACED
IN A SEPARATE FILE AVAILABLE FOR DOWNLOAD ALONG WITH THIS\end{center}

\caption{\label{Matrix-withU-50}Including coulomb interactions: Surface and
Contour plots of the WDF at=50fs for the Initial WDF of (a) zero,
(b) PseudoFermi, (c) Central}
\end{figure}
\begin{figure}[H]
\begin{center}DUE TO ARXIV SIZE CONSIDERATIONS THESE GRAPHS ARE PLACED
IN A SEPARATE FILE AVAILABLE FOR DOWNLOAD ALONG WITH THIS\end{center}

\caption{\label{Matrix-withU-100}Including coulomb interactions: Surface
and Contour plots of the WDF at=100fs for the Initial WDF of (a) zero,
(b) PseudoFermi, (c) Central}
\end{figure}
\begin{figure}[H]
\begin{center}DUE TO ARXIV SIZE CONSIDERATIONS THESE GRAPHS ARE PLACED
IN A SEPARATE FILE AVAILABLE FOR DOWNLOAD ALONG WITH THIS\end{center}

\caption{\label{Matrix-withU-200}Including coulomb interactions: Surface
and Contour plots of the WDF at=200fs for the Initial WDF of (a) zero,
(b) PseudoFermi, (c) Central}
\end{figure}
\begin{figure}[H]
\begin{center}DUE TO ARXIV SIZE CONSIDERATIONS THESE GRAPHS ARE PLACED
IN A SEPARATE FILE AVAILABLE FOR DOWNLOAD ALONG WITH THIS\end{center}

\caption{\label{Matrix-withU-1000}Including coulomb interactions: Surface
and Contour plots of the WDF at=1000fs for the Initial WDF of (a)
zero, (b) PseudoFermi, (c) Central}
\end{figure}
\begin{figure}[H]
\begin{center}DUE TO ARXIV SIZE CONSIDERATIONS THESE GRAPHS ARE PLACED
IN A SEPARATE FILE AVAILABLE FOR DOWNLOAD ALONG WITH THIS\end{center}

\caption{\label{Matrix-withU-2000}Including coulomb interactions: Surface
and Contour plots of the WDF at=2000fs for the Initial WDF of (a)
zero, (b) PseudoFermi, (c) Central}
\end{figure}

\section{Conclusions\label{sec:Conclusions}}

We have shown here a a 2D drift/scattering and 3D potential form of
the Wigner function transport equation for the case of a cylindrical
device. This is an important step in Wigner function simulations of
electronic transport since previous simulations have been restricted
to 1D drift/scattering and 1D potential. Any comparison to a real
device must answer the question of how important both radial drift/scattering
and a non 1D Coulombic charge density are. We are now prepared to
examine these questions.

The computational hurdles of solving a 2D WFE have been identified:
(1) Not all the matrices are sparse enough to fit into the RAM of
present day computers, and (2) Direct integration becomes more time
consuming as the number of simultaneous equations to solve grows large.
Some solutions of these hurdles have been described and a workable
way of numerically simulating a RTS that exhibits cylindrical symmetry
has been given. If one can separate the operator $\Omega=\mathbf{T}_{z}+\mathbf{U}+\mathbf{S}$
into the product of a drift/scattering operator and a potential operator,
then each matrix can be made sparse (overcoming the first hurdle)
and/or block diagonal (the second hurdle). This also allows the separation
of the transport problem into a two step problem, drift/scatter and
potential computations. An additional time saving method, based on
this split, is introduced. Finally, but taking advantage of the Fourier
transform nature of the potential term, one can decrease the operator
size substantially further. 

Time evolution simulations based on these method were then presented
for three different cases. Each case lead to numerical results consistent
with expectations. To the author's knowledge, this is the first proof-of-principle
of a 1D space, 2D momentum simulation. At the present time, a full
transient treatment of a forward/backwards bias sweep would take upwards
of 3 months. The work shown here still must be numerically scrutinized
in order to shorten the computer times involved. We are beginning
to work with a group of numerical experts to enhance the PDE solvers
shown here. One important item to remember is that the computational
hardware is still following Moore's law, that is, approximately doubling
in speed every 18 months. In a few years, the techniques shown here,
which push current technology to the limit, will prove to be even
more feasible in the foreseeable future.

It is our belief that the methods outlined in this paper will finally
allow a full treatment of the RTS transport problem.

\begin{acknowledgments}
The authors would like to thank Peiji Zhao and Harold Grubin for many
insightful discussions regarding RTS Wigner function simulations,
C.T. Kelley for showing us different numerical techniques for solving
the WFE, and M. Burcin Unlu for discussions on the 2D derivation.
This work has been supported but the Defense University Research Initiative
on Nanotechnology (DURINT) project grant from the Army Research Office.
\end{acknowledgments}
\bibliographystyle{apsrev}
\bibliography{greg}

\end{document}